\documentclass[acmsmall,screen,nonacm]{acmart}
\AtBeginDocument{%
  }

\usepackage{color, colortbl, xcolor}
\usepackage{url}
\usepackage{subcaption}
\usepackage{textcomp}
\usepackage{soul}
\usepackage{multirow}
\usepackage{enumitem}
\usepackage{mathtools}
\usepackage{siunitx}
\usepackage{array}
\usepackage{colortbl}
\usepackage{hhline}

\usepackage{booktabs} 
\usepackage{array}
\usepackage{xcolor}

\newcommand*{\rowstyle}[1]{
  \gdef\@rowstyle{#1}%
  \@rowstyle\ignorespaces%
}

\newcolumntype{=}{
  >{\gdef\@rowstyle{}}%
}

\newcolumntype{+}{
  >{\@rowstyle}%
}

\usepackage{arydshln}
\definecolor{LightGray}{gray}{0.97}
\definecolor{linkColor}{RGB}{6,125,233}
\definecolor{green}{rgb}{0.0, 0.65, 0.31}
\definecolor{bleudefrance}{rgb}{0.19, 0.55, 0.91}
\definecolor{ceruleanblue}{rgb}{0.16, 0.32, 0.75}
\definecolor{grey}{HTML}{969696}
\definecolor{violet}{HTML}{756bb1}
\definecolor{dgrey}{HTML}{01665e}
\definecolor{lgrey}{HTML}{5ab4ac}
\definecolor{dgreen}{HTML}{005a32}
\definecolor{purple}{HTML}{ae017e}

\definecolor{editCol}{HTML}{000000}
\definecolor{maskCol}{HTML}{c51b7d}
\definecolor{lrColor}{HTML}{8856a7}
\definecolor{trColor}{HTML}{d01c8b}
\definecolor{ctColor}{HTML}{4dac26}
\definecolor{brickred}{HTML}{f03b20}
\definecolor{improveCol}{HTML}{253494}
\definecolor{worsenCol}{HTML}{d7191c}
\definecolor{DarkBlue}{HTML}{00008B}
\definecolor{mscolor}{HTML}{01665e}
\definecolor{nmscolor}{HTML}{bf812d}
\definecolor{lgreen}{HTML}{ccece6}
\definecolor{dolive}{HTML}{308014}

\definecolor{editCol}{HTML}{000000}
\definecolor{maskCol}{HTML}{c51b7d}
\definecolor{lrColor}{HTML}{8856a7}
\definecolor{trColor}{HTML}{d01c8b}
\definecolor{ctColor}{HTML}{4dac26}
\definecolor{brickred}{HTML}{f03b20}
\definecolor{improveCol}{HTML}{253494}
\definecolor{worsenCol}{HTML}{d7191c}
\definecolor{lgreen}{HTML}{e0f3db}
\definecolor{dpink}{HTML}{CD1076}
\definecolor{pink}{HTML}{FED2D2}
\definecolor{soothinggreen}{HTML}{4dac26}
\definecolor{darkred}{HTML}{8B0000}

\definecolor{dblue}{HTML}{104E8B}
\definecolor{violet}{HTML}{8A2BE2}
\definecolor{mscolor}{HTML}{01665e}
\definecolor{nmscolor}{HTML}{d8b365}
\definecolor{deepgrey}{HTML}{525252}
\definecolor{dslate}{HTML}{2F4F4F}
\definecolor{dolive}{HTML}{556B2F}
\definecolor{teal}{HTML}{388E8E}
\definecolor{mscolor}{HTML}{01665e}
\definecolor{nmscolor}{HTML}{d8b365}

\definecolor{aicolor}{HTML}{018571}
\definecolor{occolor}{HTML}{ff7799}

\definecolor{srcolor}{HTML}{e34a33}
\definecolor{smcolor}{HTML}{253494}
\definecolor{srsmcolor}{HTML}{7fcdbb}
\definecolor{bothcolor}{HTML}{fe9929}
\definecolor{onecolor}{HTML}{018571}
\definecolor{marroon}{HTML}{881c1c}

\colorlet{tablerowcolor4}{gray!50} 

\newcommand*{\textlabel}[2]{%
  \edef\@currentlabel{#1}
  \phantomsection
  #1\label{#2}
}
\usepackage{tcolorbox}

\colorlet{tableheadcolor}{gray!25} 
\colorlet{tablerowcolor}{gray!15} 
\colorlet{tablerowcolor2}{gray!45} 
\colorlet{tablerowcolor3}{gray!25} 

\newcommand{\rowcolmedium}{\rowcolor{tablerowcolor2}}
\newif{\ifhidecomments}
  \hidecommentsfalse 
\ifhidecomments
    \newcommand{\veda}[1]{}
    \newcommand{\andy}[1]{}
    \newcommand{\jash}[1]{}
    \newcommand{\haylee}[1]{}
    \newcommand{\ziang}[1]{}
    \newcommand{\vedant}[1]{}
    \newcommand{\koustuv}[1]{}
\else
    \newcommand{\veda}[1]{\textbf{\small\sffamily{\textcolor{DarkBlue}{[#1 -- Veda]}}}}
    \newcommand{\andy}[1]{\textbf{\small\sffamily{\textcolor{dgreen}{[#1 -- Andy]}}}}
    \newcommand{\jash}[1]{\textbf{\small\sffamily{\textcolor{dolive}{[#1 -- Jash]}}}}
    \newcommand{\haylee}[1]{\textbf{\small\sffamily{\textcolor{violet}{[#1 -- Haylee]}}}}
    \newcommand{\ziang}[1]{\textbf{\small\sffamily{\textcolor{brickred}{[#1 -- Ziang]}}}}
    \newcommand{\vedant}[1]{\textbf{\small\sffamily{\textcolor{soothinggreen}{[#1 -- Vedant]}}}}
    \newcommand{\koustuv}[1]{\textbf{\small\sffamily{\textcolor{dpink}{[#1 -- Koustuv]}}}}
  \fi

\newcommand{\trc}{\texttt{Trucey}}
\newcommand{\cgpt}{\texttt{Control-AI}}
\newcommand{\hbk}{\texttt{Control-NoAI}}

\renewcommand{\textrightarrow}{$\rightarrow$}

\colorlet{tableheadcolor}{gray!25} 
\colorlet{tablerowcolor}{gray!5} 

\definecolor{neutralCol}{HTML}{dd1c77}
\definecolor{neutralGreen}{HTML}{31a354}
\definecolor{NewBlue}{HTML}{1879ba}
\definecolor{bleudefrance}{rgb}{0.19, 0.55, 0.91}  
\definecolor{AfTrColor}{HTML}{0868ac}  
\definecolor{BfTrColor}{HTML}{a8ddb5}  

\definecolor{AfCtColor}{HTML}{b10026}  
\definecolor{BfCtColor}{HTML}{fd8d3c}

\graphicspath{ {figures/} }

\newcommand{\para}[1]{\vspace{0.5em}\noindent\textbf{#1}~}

\setcopyright{acmlicensed}
\copyrightyear{2026}
\acmYear{2026}
\acmDOI{XXXXXXX.XXXXXXX}
\acmConference[Conference acronym 'XX]{Make sure to enter the correct
  conference title from your rights confirmation email}{June 03--05,
  2018}{Woodstock, NY}
\acmISBN{978-1-4503-XXXX-X/2018/06}

\begin{document}

\title[AI-Mediated Negotiation: Design Reflections and Lessons]{AI-Mediated Negotiation: Design Reflections and Lessons}

\author{Veda Duddu}
\email{vduddu2@illinois.edu}
\orcid{0009-0001-6443-6239}
\affiliation{
    \institution{University of Illinois Urbana Champaign}
    \city{Urbana}
    \state{Illinois}
    \country{USA}
}
\author{Jash Rajesh Parekh}
\orcid{0000-0003-3310-4634}
\affiliation{%
 \institution{University of Illinois Urbana-Champaign}
 \city{Urbana}
 \state{IL}
 \country{USA}}
\email{jashrp2@illinois.edu}

\author{Andy Mao}
\orcid{0009-0007-6060-9730}
\affiliation{%
 \institution{University of Illinois Urbana-Champaign}
 \city{Urbana}
 \state{IL}
 \country{USA}}
\email{hanqim2@illinois.edu}

\author{Hanyi Min}
\orcid{0000-0002-0095-8513}
\affiliation{%
 \institution{University of Illinois Urbana-Champaign}
 \city{Urbana}
 \state{IL}
 \country{USA}}
 \email{hanyimin@illinois.edu}

\author{Ziang Xiao}
\orcid{0000-0003-3368-0180}
\affiliation{%
 \institution{Johns Hopkins University}
 \city{Baltimore}
 \state{MD}
 \country{USA}}
 \email{ziang.xiao@jhu.edu}

\author{Vedant Das Swain}
\orcid{0000-0001-6871-3523}
\affiliation{%
 \institution{New York University}
 \city{New York City}
 \state{NY}
 \country{USA}}
 \email{v.das.swain@nyu.edu}

\author{Koustuv Saha}
\orcid{0000-0002-8872-2934}
\affiliation{%
 \institution{University of Illinois Urbana-Champaign}
 \city{Urbana}
 \state{IL}
 \country{USA}}
\email{ksaha2@illinois.edu}

\renewcommand{\shortauthors}{Duddu et al.}

\begin{abstract}
Conversational AI promises a new kind of preparation for high-stakes workplace negotiations---personalized, interactive, and capable of simulating realistic resistance. That promise is intuitive. We built \trc{}, a theory-driven coaching system to test it. The system encoded four assumptions: that articulation supports clarification, that personalization builds strategic competence, that chunked delivery reduces cognitive load, and that structured scaffolding removes metacognitive burden. A pre-registered experiment (N=267) and interviews (N=15) complicated each of them. Notably, the static handbook we included as a passive control outperformed both AI conditions on empowerment and usability. We reflect on why: each assumption encoded a specific model of how preparation unfolds, and the findings revealed that conversational AI imposes a linear execution model on a task that is fundamentally recursive. We identify an unexamined scope condition on established HAI design guidelines and close with a sequencing principle---map before path, path before simulation---for future AI coaching design.
\end{abstract}

\ccsdesc[500]{Human-centered computing~Empirical studies in HCI}

\keywords{workplace negotiation, conversational AI, design reflection, cognitive load, human-AI interaction, AI coaching}


\maketitle
\section{Introduction}
Workplace negotiations---requesting a promotion, pushing back on a decision, asserting a boundary---are rarely just strategic challenges. For most workers, the barrier is psychological: fear of retaliation, uncertainty regarding the unfolding of the conversation, and a pervasive sense of power asymmetry that renders the act of initiating risky~\cite{Bradley_Campbell_2016, brett2016negotiation}. Conversational AI has emerged as a promising approach for addressing these barriers, offering personalized, interactive preparation that static resources cannot provide.
\trc{} is a theory-driven AI coaching system we built to lower these barriers, operationalizing negotiation frameworks through situational calibration, role-based rehearsal, and structured scaffolding.

In this paper, we share our design reflections and lessons learned from building AI-mediated coaching tools for high-stakes, preparatory workplace tasks. We built \trc{} around four assumptions about how workers prepare under stress, each grounded in established theory. Our empirical study---a pre-registered between-subjects experiment (N = 267) and semi-structured interviews (N = 15)---complicated those assumptions in ways we did not anticipate. Crucially, the static handbook, \hbk{}, we included as a passive control, outperformed both AI conditions on psychological empowerment and usability. 

That result sent us back to our design assumptions. What we found was not that \trc{} failed; rather, the conversational delivery mechanism was structurally mismatched with the cognitive demands of the task. 
Conversational AI encodes specific assumptions about how preparation should unfold. 
We highlight that those assumptions are not neutral---and when they misalign with the task's cognitive demands, they do not just fail to help; they actively get in the way.

We describe our design rationale and assumptions (Section~\ref{sec:design}), briefly overview the study and its findings (Sections~\ref{sec:study}--\ref{sec:findings}), reflect on where those assumptions broke and why (Section~\ref{sec:reflection}), and close with scope conditions and forward-looking design directions (Sections~\ref{sec:scope}--\ref{sec:directions}).

\section{Why We Designed \trc{} This Way}
\label{sec:design}
\trc{} was designed around four theoretically grounded assumptions about how workers prepare for high-stakes negotiations under stress. Each mechanism instantiates a specific model of preparation drawn from negotiation theory, cognitive science, and human–AI interaction.

\para{Mechanism 1:  Situational Calibration.} We assumed that articulating one's negotiation context supports preparation by structuring and externalizing understanding. Workplace negotiation strategies are shaped by relational history, power asymmetries, and institutional context~\cite{brett2016negotiation}. \trc{} elicits descriptions of the user’s situation---including role, prior interactions, and supervisor characteristics---and conditions subsequent guidance on this context. This design reflects the view that users are active participants in constructing their understanding through articulation~\cite{fischer1998beyond}.

\para{Mechanism 2: Role-Based Simulation.} Our second assumption held that exposure to realistic, context-specific pushback supports fear reduction and builds confidence. Avoidance of negotiation is often driven by anticipated resistance without opportunities for safe exposure. Prior work suggests that engaging with challenging, anxiety-inducing scenarios can support fear processing and self-efficacy~\cite{foa1986emotional, bandura1977self}. \trc{} simulates a supervisor whose responses reflect the user's described context and communication style, creating a rehearsal environment that approximates workplace dynamics beyond generic, agreeable AI interactions~\cite{sharma2023cognitive,shaikh2024rehearsal,dasswain2025ai}.

\para{Mechanism 3: Contextual Layering.} A third assumption guided this mechanism: that incremental, chunked delivery reduces cognitive load during preparation under stress. Preparation under stress constrains attention and information processing. Cognitive load theory suggests that managing information complexity supports learning~\cite{sweller2011cognitive}, particularly in high-stress contexts~\cite{porcelli2017stress}. \trc{} delivers guidance incrementally across conversational turns, aligning with human–AI interaction guidelines on staged and interpretable system behavior.

\para{Mechanism 4: Iterative Response Alignment.} Finally, we assumed that structured scaffolding reduces metacognitive burden by managing the progression of preparation. Users’ needs evolve during preparation, requiring ongoing adjustment of tone, strategy, and challenge level. \trc{} embeds structured feedback loops within the interaction, enabling real-time recalibration. This reflects models of adaptive conversation management~\cite{Bradley_Campbell_2016} and prior work on reducing cognitive burden in AI-assisted tasks~\cite{buccinca2021trust}. The mechanism also draws on perspectives that locate empowerment in perceived control over task progression~\cite{spreitzer1995psychological}. \trc{} provided no structured overview of the preparation space --- while prior turns remained visible by scrolling, a linear transcript does not substitute for a navigable map of what has been covered and what remains.

Across these mechanisms, \trc{} assumes that preparation is supported through a combination of articulation, guided feedback, and rehearsal within a structured interaction. Rather than relying on static resources, the system positions preparation as an interactive process in which users engage with context-sensitive guidance and simulated experience. The expectation is that this combination supports both affective readiness and strategic competence. 
\section{Study Overview and Findings}
\subsection{Study Overview}
\label{sec:study}
We conducted a pre-registered between-subjects experiment (N=267) comparing three conditions: \trc{} (AI with theory-driven scaffolding), a generic LLM baseline (\cgpt{}), and a static theory-grounded handbook (\hbk{}). By separating theoretical scaffolding from conversational interactivity, this design allowed us to ask whether each ingredient was actually pulling its weight. We complemented quantitative outcomes---fear, empowerment, usability---with semi-structured interviews (N=15) to understand the experiences driving them.

\subsection{Findings}
\label{sec:findings}
The results did not go the way we expected. \trc{} reduced fear more effectively than the \cgpt{} ($d = -0.27, p<.05$)---that part held. However, the \hbk{}, our passive control, outperformed both AI conditions on psychological empowerment $(d = -0.40, p<.01)$ and usability $(80.47 vs. 74.22, d = -0.32, p<.05)$. Interactivity did not translate into greater perceived competence. If anything, it got in the way.
The interviews contextualized these results~\cite{duddu2025does}. Participants sought a structural overview before engaging with specific guidance — a sequence \trc{}'s turn-by-turn delivery systematically inverted. That inversion, and what it revealed about our design assumptions, is what this poster is about.
This paper focuses on design reflection rather than findings reporting; further methodological and empirical details can be found in prior work~\cite{duddu2025does,duddu2026not}.

\section{Where the Design Assumptions Broke}
\label{sec:reflection}
We built \trc{} around assumptions we considered well-grounded. Each was theoretically justified. Each broke in ways that taught us something we could not have seen from the theory alone (Table~\ref{tab:assumption-matrix}).

\begin{table}[t!]
\centering
\sffamily
\footnotesize
\caption{Assumption/boundary matrix: four design mechanisms, what each assumes, and where each breaks.}
\label{tab:assumption-matrix}
\begin{tabular}{p{1.6cm} p{3.4cm} p{3.4cm} p{3.4cm}}
\textbf{Mechanism} & \textbf{Assumption Encoded} & \textbf{Holds When} & \textbf{Breaks When} \\
\toprule
\rowcolmedium \multicolumn{4}{l}{\textit{Situational Calibration}}\\
~ & Articulating context structures and externalizes understanding — conversation functions as self-clarification. & Users are elaborating on something they partially understand and need to surface — articulation genuinely moves their thinking forward. & Users already know what they know — the task requires recursive revision, not description. Conversational AI solicits but cannot scaffold iterative refinement. \\
\rowcolmedium \multicolumn{4}{l}{\textit{Role-Based Simulation}}\\
~ & Exposure to realistic, context-specific pushback reduces fear and builds confidence through rehearsal. & Users accept the simulation as sufficiently realistic — the affective engagement is credible enough to approximate actual stakes. & Users question whether AI can reproduce the embodied unpredictability of real workplace power dynamics — authenticity breaks down and fear processing does not transfer. \\
\rowcolmedium \multicolumn{4}{l}{\textit{Contextual Layering}}\\
~ & Incremental chunked delivery reduces cognitive load by pacing information across turns. & Task is sequential — each chunk is complete without reference to the whole, and users do not need to navigate back or assess coverage. & Task is recursive — users need positional awareness, the ability to revisit earlier content, and a persistent overview of what remains. Chunking without overview increases rather than reduces load. \\
\rowcolmedium \multicolumn{4}{l}{\textit{Iterative Response Alignment}}\\
~ & Structured scaffolding reduces metacognitive burden by managing the progression of preparation. & The interaction model matches the task's cognitive structure — users just need to respond, not navigate. & Preparation is recursive and users need to maintain continuity across turns — the conversation holds no persistent state, so tracking prior context shifts entirely onto the user. Invisible structure produces anxiety rather than relief. \\
\bottomrule
\end{tabular}
\Description[table]{This table presents a four-row assumption/boundary matrix for Trucey's design mechanisms. Each row covers one mechanism — Situational Calibration, Role-Based Simulation, Contextual Layering, and Iterative Response Alignment — and traces the assumption encoded in its design, the conditions under which that assumption holds, and the conditions under which it breaks. The matrix shows that three of the four mechanisms fail under recursive, emotionally loaded preparation tasks, and that the failures are structural rather than incidental.}
\end{table}

\para{Articulation elicited retrieval, not clarification.} We assumed that prompting workers to describe their negotiation context would function as self-clarification—helping them structure and externalize their thinking. The interaction however elicited knowledge telling; users retrieved and expressed what they already knew rather than engaging in the recursive restructuring necessary for preparation \cite{cress2023co}. We conflated articulation with clarification. While we successfully prompted users to describe their negotiation context, we failed to provide the recursive space needed for them to refine that thinking. The conversational model solicited description, but it offered no way to revisit, challenge, or evolve those initial thoughts.

\para{Personalization shaped rehearsal affect, not strategic competence.} We assumed that grounding coaching in supervisor-specific context would produce more targeted, actionable guidance than generic advice. On strategic outcomes, this did not hold---empowerment gains between \trc{} and our control condition were negligible. However, we found that personalization did matter for affect: it made the simulated supervisor feel more realistic, which directly reduced user fear \cite{foa1986emotional, bandura1977self}. We learned that we had built a single mechanism expecting it to do two things, though it could only serve an affective rather than a cognitive purpose.

\para{Chunking without overview increased cognitive load.} Following established HAI design guidelines \cite{amershi2019guidelines} in good faith, we utilized incremental, chunked delivery. Our results surfaced an unexamined scope condition on these guidelines: chunking reduces cognitive load when each unit can be processed independently and the task unfolds sequentially. Negotiation preparation is fundamentally recursive---users move between concerns, revisit earlier decisions, and reassess their progress. Because \trc{} chunked information without a persistent overview, users lacked positional awareness, unable to navigate back or gauge their completeness. While the guidelines were derived from sequential task-completion contexts, they inadvertently increased load when applied to recursive, non-linear domains.

\para{Scaffolding imposed a linear execution model on a recursive task.} We assumed that theory-driven scaffolding would reduce metacognitive burden by managing the progression of preparation. Instead, under time pressure, the invisible structure produced uncertainty: because users could not see the ``map'' of the conversation, they could not anticipate what remained or assess if they had covered everything. Crucially, the interaction model forced a linear progression on a recursive task. Users were required to hold prior insights in working memory and reintroduce them manually---the conversation had no persistent state they could return to. This shifted the burden of maintaining cognitive continuity entirely onto the user, transforming the scaffolding from a support mechanism into a source of coordination effort.

\para{Guided delivery hindered psychological ownership.} Our competence-building model assumed that receiving personalized, expert-calibrated feedback would build applicable understanding. However, we underestimated the role of ownership in this system. Psychological ownership of knowledge develops through control over it and intimate familiarity with it \cite{pierce2001toward}. \trc{}’s guided delivery denied both; users followed the system's progression and saw only what the system surfaced. In contrast, the \hbk{} allowed users to navigate, skim, and return to content—granting them the control necessary to ``own'' the knowledge. We now believe that control over the organization of information is not incidental to empowerment; it is constitutive of it.

These break points converge on two principles that extend beyond \trc{}. First, conversational AI systems without persistent state or navigable overview are poorly suited to tasks where understanding is recursive rather than sequential; the medium externalizes navigation but internalizes state, forcing users to maintain cognitive continuity that the interaction model cannot preserve. Second, chunked incremental delivery reduces cognitive load when task structure is sequential, but it increases cognitive load when task structure is recursive and users require persistent positional awareness. Together, these findings identify an unexamined scope condition on established HAI design guidelines: they are largely derived from sequential task-completion paradigms. These limitations were invisible until we introduced a control condition that made the contrast legible, demonstrating that the failure of current scaffolding is not just an implementation detail but a fundamental mismatch between interaction design and task structure.

We did not measure cognitive load or navigation effort directly; these claims are inferred from empowerment outcomes and interview accounts.

\section{Reflection and Design Directions}
\subsection{Reflections on Generalizability and Scope}
\label{sec:scope}
Our evaluation suggests that the efficacy of conversational AI in this domain is highly dependent on task structure. 
\trc{}'s rehearsal mechanism proved effective; it reduced negotiation-related fear more consistently than the unstructured baseline. This success likely stems from the fact that simulation is inherently sequential---a workplace negotiation unfolds turn-by-turn, making a conversational medium well-suited to the task. When the medium’s interaction model aligns with the task’s temporal structure, the design succeeds.

However, we are cautious about the generalizability of these findings. 
Our qualitative sample (N=15) was disproportionately comprised of early-career professionals in technology. 
The ``map-before-path'' preference we observed may reflect the systematic, structured preparation styles common in these cohorts as much as it reflects a universal cognitive requirement. Furthermore, our theoretical framing relies on cognitive and educational frameworks---such as knowledge transformation and global-before-local learning---which are not direct empirical equivalents to high-stakes workplace preparation. We treat these theories as interpretative lenses that illuminate the observed patterns rather than as fixed empirical claims regarding negotiation behavior.
The general principle we posit is this: conversational AI is effective when the medium's linear interaction model matches the task's cognitive structure. Where that match breaks down, the system imposes a hidden cognitive tax on the user—one that current interaction models are not designed to mitigate. We cannot fully disentangle interactivity from navigability as drivers of the empowerment gap---a limitation future work should aim to address through matched-format designs.

Beyond the specific findings, this study offers a cautionary lesson for the HCI and CSCW community. Conversational AI is increasingly substituting for structured static resources across domains~\cite{das2024teacher,dasswain2025ai}. Our findings suggest the costs of that substitution are not obvious. In fact, even a well-designed static resource can outperform a theory-driven AI system on the outcomes that matter most. Researchers and designers building AI for high-stakes preparatory contexts should expect the same tension: the medium encodes assumptions about how preparation unfolds, and those assumptions may not match how users actually prepare under stress.

\subsection{Forward-Looking Design Directions}
\label{sec:directions}
If we were to build this again, we would not change the components---we would change the order.

\para{Map before path.} Users need structural orientation before engagement. A static, navigable, exportable overview of the full preparation space---organized by theoretical framework, covering strategy, relationship dynamics, and likely resistance---should come first. Not as a preamble to the AI, but as the primary resource. Users should be able to see the full scope of what preparation requires, engage with different dimensions in any order, and return to earlier material as their understanding develops.

\para{Path before simulation.} Conversational AI earns its place once users have the map and have identified where they need depth. At that point, dialogue is no longer imposing a sequence---it is serving one the user has already chosen. The system becomes a tool users direct rather than a guide users follow.

\para{Simulation last.} Rehearsal belongs here because it is inherently sequential---it practices something that is itself linear. Once users have strategic orientation and have selected what they want to work on, structured roleplay with calibrated resistance is the right next step.

This sequencing principle extends beyond negotiation. Career coaching, legal preparation, medical decision-making, performance reviews---any high-stakes context requiring both strategic overview and emotional rehearsal faces the same underlying tension. The medium is not neutral. Matching interaction architecture to cognitive task structure is not a refinement of AI coaching design. It is a precondition for it.

\section*{AI Disclosure}
The authors used AI-assisted writing tools to support grammar and prose editing. All intellectual content, arguments, and analytical claims are the authors' own.

\bibliographystyle{ACM-Reference-Format}
\bibliography{sections/0references}
\end{document}
\endinput